# Protocols for Bio-Inspired Resource Discovery and Erasure Coded Replication in P2P Networks

Sabu M. Thampi[†], Chandra Sekaran K[††]
[†]Rajagiri School of Engineering and Technology, Kochi, Kerala, India
[††]National Institute of Technology Karnataka, Surathkal, Karnataka, India
*sabum@rajagiritech.ac.in, kch@nitk.ac.in*

*Abstract*: Efficient resource discovery and availability improvement are very important issues in unstructured P2P networks. In this paper, a bio-inspired resource discovery scheme inspired by the principle of elephants migration is proposed. A replication scheme based on Q-learning and erasure codes is also introduced. Simulation results show that the proposed schemes significantly increases query success rate and availability, and reduces the network traffic as the resources are effectively distributed to well-performing nodes.

*Keywords*: Unstructured P2P network; Searching; Replication; Q-learning; Bio-inspired; Elephants migration.

## I. Introduction

P2P networks are a recent addition to the already large number of distributed system models. P2P networking has spawned immense attention worldwide amongst both Internet users and computer professionals. P2P computing takes advantage of existing computing power, computer storage and networking connectivity, allowing users to leverage their collective power to the benefit of all.

There are mainly three different architectures for P2P systems: *centralized, decentralized structured and decentralized unstructured*. In the centralized model, such as Napster [2], central index servers are used to maintain a directory of shared files stored on peers with the intention that a peer can search for the location of a desired content from an index server. On the other hand, this design makes a single point failure and its centralized nature of the service creates systems susceptible to denial of service attacks. Decentralized P2P systems have the advantages of eliminating dependence on central servers and providing freedom for participating users to swap information and services directly between each other. In decentralized structured models, such as Chord [3], Pastry [4], and CAN [5], the shared data placement and topology characteristics of the network are strongly controlled on the basis of distributed hash functions. In decentralized unstructured P2P systems, such as Gnutella [6] and KaZaA [7], there is neither a centralized index nor any strict control over the network topology or file placement. Nodes joining the network, following some loose rules, form the network. The resulting topology has certain properties, though the placement of objects is not based on any knowledge of the topology [8]. The decentralization makes available the opportunity to utilize unused bandwidth, storage and processing power at the periphery of the network. It diminishes the cost of system ownership and maintenance and perks up the scalability.

During a search operation in a P2P system, several query packets pass through the network searching for the target objects. The heterogeneity of these query packets creates a local traffic disparity and congestion. The downloading of large objects in response to requests also causes congestion in nodes. One proficient method for forestalling this load concentration is providing redundant copies of objects into various sites. This is called replication which is a widely accepted technique in distributed environment, where data is stored at more than one site for performance and reliability reasons. Replication increases object availability and fault tolerance. Single node failures, like crashes of nodes, can be tolerated as faults within the system as a whole facilitated with the help of the redundancy introduced by replicas. Replicating objects to multiple sites has several issues such as selection of objects for replication, the granularity (size) of replicas, and choosing appropriate site for hosting new replica [9, 10].

This paper proposes techniques for efficient resource discovery and availability improvement. The resource discovery scheme employs a bio-inspired concept for discovering desired objects and spreading popular objects in the network efficiently. The Reed-Solomon erasure code [11] is employed to offer erasure coded replication of popular objects into well-performing peers in the P2P network. In the proposed replication scheme 'Q-Erasure', a node in the network first selects a popular object for performing erasure coded replication. The selected object is divided into 'k' fragments and these 'k' fragments are used to create 'n' (n≥k) erasure coded blocks which will be replicated to suitable sites. The original object can be reconstructed from 'n' erasure coded 'k' blocks. Thus, the process of replication involves selection of suitable objects, creation of erasure coded blocks to distribute among several sites, selection of appropriate sites for hosting erasure coded bocks and replication of selected objects to chosen sites. The entire process is based on Q-learning [45].

The rest of the paper is organized as follows. Section II briefs the related work on resource discovery and replication in P2P networks. Section III discusses the process of elephants migration. Section IV describes the proposed bio-inspired resource discovery scheme for decentralized unstructured P2P networks. Section V introduces the Q-learning based replication scheme. The simulation environment is discussed in section VI. Section VII presents the results and a discussion on them. Finally, section VIII concludes the paper.

## II. Related Work

This section briefly reviews the previous work on resource discovery and replication in P2P networks.

*A) Bio-Inspired Approaches*

The search schemes for unstructured P2P network are generally classified as *blind* [8, 15, 16, 17] and *informed* [18, 19]. Most of the search techniques are either random techniques or probabilistic schemes. In a blind search, nodes do not keep information about object location. In an informed search, nodes gather some metadata that assist the search operation. Bio-inspired approaches are recently introduced in a variety of ways for different applications such as resource discovery.

A hybrid ant-inspired search algorithm (HASA) for P2P media streaming distribution in Ad Hoc networks is proposed in [20]. It utilizes the merits of random walkers and ant-inspired algorithms for search in unstructured P2P networks, such as low transmitting latency and less redundant query messages.

In [21] focuses on the free-rider problem in unstructured P2P networks, and proposes a new search algorithm, called "AntSearch", to reduce the redundant messages during a query flooding. Each peer maintains a pheromone value to present its success rate of past processed queries, and keeps a list of pheromone values of its immediate neighbors. The main idea of the AntSearch is using pheromone values to identify the free-riders, prevent sending messages to those peer in order to reduce the redundant messages.

A Swarm Intelligence Technique Bees Algorithm called P2PBA (Peer to Peer file sharing - Bees Algorithm) which is based on the lines of food search behavior of Honey Bees is proposed in [22]. The scheme optimizes the search process by selectively going to more promising honey sources and scan through a sizeable area.

In [23] proposes a P2P network based location search algorithm which can be used to establish connections in Internet Telephony. By using the location search algorithm, the caller can identify the peer it is calling. The algorithm is based on the notion of *gradient search* and is applicable to unstructured networks. It is inspired by a biological phenomenon called *haptotaxis*. The algorithm performs at par with DHT-based location search algorithms.

The search algorithm proposed in [24, 25] is termed ImmuneSearch which draws its basic inspiration from natural immune systems. It is implemented separately by each individual peer participating in the network and is completely decentralized in nature. Instead of flooding query messages, ImmuneSearch uses an immune system. The topology evolution coupled with proliferation and mutation help the P2P network to develop 'memory'. Due to this, the search efficiency of the network improves as several individual peers perform search. In [26] presents SemAnt, an algorithm for distributed query routing based on the Ant Colony Optimization meta-heuristic.

The techniques discussed in [27] transform discrete sequence equations into new P2P protocols called *sequence protocol*. The methodology defines a set of procedures that takes as input any instance of the targeted class of models, and produces an output that is a P2P protocol. The protocol can be derived systematically from multi-variable sequence equations, which are perhaps originally based on natural phenomena. Sequence protocols are self-adaptive, scalable, and fault-tolerant, with applicability in P2P settings like Grids.

In [28] presents an algorithm designed for performing autonomous careful dissemination of messages within a network. It constitutes the communication layer called 'Personal Intelligent Agent Framework' which is intended to help users transparently share information. The algorithm works in a fully decentralized way, using epidemic diffusion mechanism and artificial ants paradigm to achieve self-organization and information flows management. In [29] proposes a bio-inspired cache replacement algorithm that considers the balance between supply and demand for media streams based on the division of labor and task allocation. In this scheme, a peer estimates the supply and demand based on locally available and passively obtained information.

In [30] presents 'Self-Chord', a bio-inspired P2P algorithm that can be profitably adopted to build the information service of distributed systems, in particular Computational Grids and Clouds. In Self-Chord, a set of ant-inspired mobile agents move and reorder the resource keys in a ring of peers in a self-organizing fashion, without any predetermined association between keys and peers. The keys are fairly distributed over the peers, thus improving the balancing of storage responsibilities. Another advantage is the reduction in the maintenance load due to the reorganization of the keys by mobile agents in logarithmic time as new peers join the ring.

*B) Erasure Coded Replication Schemes*

Several solutions are proposed by researchers to increase the performance of replication process in P2P systems [8, 34-39]. In [40] P2P replication techniques are classified into three categories based on size of files: *full file replication* in which "full" files are replicated at multiple peers based upon which node downloads the file. This strategy is simple to implement. However, replicating larger files at one single file can be cumbersome in space and time [41]; *block level replication* divides each file into an ordered sequence of fixed size blocks. This is also advantageous if a single peer cannot store whole file. A limitation of block level replication is that during file downloading it is required that enough peers are available to assemble and reconstruct the whole file. Even if a single block is unavailable, the file cannot be reconstructed; *erasure codes replication* provides the capability that the original files can be constructed from less number of available blocks. The replication of small erasure coded fragments saves important resources such as storage space and bandwidth.

While our proposed replication scheme focuses on erasure codes, techniques based on erasure codes are only discussed here. A review of replication techniques are discussed in [35]. An autonomous replication technique which stands on erasure codes is proposed in [42]. This scheme uses randomized decisions extensively to tolerate autonomous peer actions. Each member of the P2P community hoards some subset of the shared files entirely on their local storage, called the member's hoard set, and pushes replicas of its hoard set to peers with excess storage using an erasure code. The weaknesses of this scheme are: it does not select objects based on their popularity; objects are randomly selected from the hoard set and replicated. The target peers for hosting replicas are also chosen randomly, there by the behavior of peers are not considered for selection.

A replication Protocol – Reperasure using Erasure code -a layer on top of the basic DHT- in P2P Storage Network is proposed in [43]. Reperasure protocol rests upon the notion that when many blocks with erasure code are stored inside, a distributed hash table (DHT) can be regarded as a super-reliable and high performance disk. The scheme offers higher availability than full replicas and achieves better network and storage utilization via parallel access to the blocks.

The decentralized replication algorithms proposed in [44] deal with storage allocation and replica placement. Three heuristic algorithms - *a random algorithm*, a *group partition algorithm* that relies on peers' forming groups and a *highest available first (HAF) algorithm*- a *greedy algorithm*, are proposed. The three replication schemes employ the erasure-coded blocks for replication. The main advantage is that the greedy algorithm achieves higher availability especially when peers share a small amount of storage space for replication and when high available peers in the system are rare. The issue of overwriting of same erasure coded block by other nodes in a peer is not addressed in any of these schemes. The peers are selected randomly without considering peer attributes such as bandwidth, degree of the node etc. Hence past performance of peers is not reckoned. The method for accommodating new erasure blocks in the case of storage exhaustion is not provided.

### III. Migration of Elephants

Elephants are incredibly social creatures that have lasting memories, and can communicate over long distances through low range sound waves. They show a range of cognitive abilities and social behavior. The African elephant is the largest living land mammal. There are two kinds of elephants: *African* and *Asian elephants*. African elephants are larger than Asian elephants. African elephants inhabit a diverse array of habitats [31]. The elephants migrate and normally follow the same migratory routes every year. The migration occurs typically at the commencement of dry season. The animals move toward more suitable locations near rivers and water sources. The environmental conditions considerably affect the migration distances. The distance being covered during migration by African elephants is more than 100 km in dry seasons. Asian elephants residing in the dense forests of southern India, travels between 20 and 50 km during migration. However, when the rainy season appears, elephant herds go back to native regions to feed on the lush green vegetation. Thus, the migration of elephants allows time for the re-growth of plants in fatigued scraping areas [32, 33].

The proposed resource discovery scheme utilizes the principles of African elephants migration during dry seasons for food. The proposed scheme classifies a node (peer) in the network as ordinary nodes and power nodes. The presence of fewer amounts of popular objects in its neighborhood makes a peer reside in the *dry region* and hence for finding out more popular resources it moves to a resource rich region called *wet region*. Due to this, success rate is improved a lot. Meanwhile, powerful nodes in the network try to increase the availability of popular objects in the dry region. This leads the nodes to shift to the previous region after the nodes satisfy with the presence of sufficient quantity of popular objects. Thus, the network load is managed effectively. For increasing the availability in the dry region and other parts of the network, a replication scheme based on erasure codes and Q-learning is proposed in this paper.

### IV. Proposed Bio-Inspired Search Scheme

The P2P system model comprises nodes and files (objects). There are a few neighboring nodes ($n_1$, $n_2$, $n_3...n_n$) associated with each node 'p'. The term 'file' stands for any general content in a node or peer. The number of neighbors connected (links) to a node is called its degree. Peers with a large number of degrees receive more number of queries than peers with a small number of degrees. This is due to high availability of objects and more number of links in high degree peers.

The topology of P2P networks is modeled as a network with an undirected graph G whose nodes represent hosts and edges represent internet connections between those hosts. Nodes are usually very dynamic, where some can join and leave the network in the order of seconds whereas other nodes stay for an unlimited period of time. When a user requests a file, a search for the file is initiated and other nodes in the network need to be queried if the file is not available locally. A query is composed of one or more words. Every query message has a message-id, and this ID is stored in a peer receiving the query. Hence, if the queries with same message–id come to the same peer another time, the query is discarded.

The random walk proposed in [8] with a few added features is employed for discovering resources in the P2P network. For a query generated in a node N, the shared folder of the node which contains the shared objects is checked for a query match. In case the desired object is found, query is dropped; else the query is forwarded to K walkers which are neighbors of N. The life of a walker is determined by the current value of Time-To-live (TTL) parameter. TTL is the number of hops to be visited during a search operation. However, each neighbor forwards a walker to only one node and from that node onwards the query is forwarded as in the previous hop until the TTL expires or desired result is found.

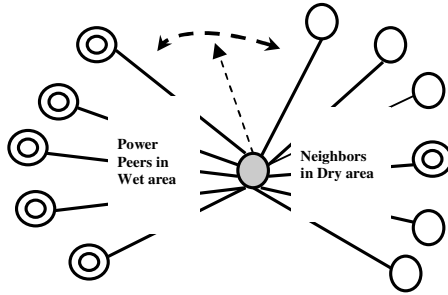

Figure 1. Dry and Wet areas in a P2P network

If the popular objects are not available in nearby nodes; the query will take maximum TTL value and thus the number of messages increase drastically creating heavy network traffic. At the same time, the success rate decreases. The user may unhappy with the performance of the system and decides to leave the system permanently. Hence, queries should be appropriately routed to well-performing nodes and a proper mechanism for increasing availability of popular objects in all categories of peers should be employed. The first issue discusses a suitable criterion for routing queries in case of shortage of popular objects in the current search path. The later demands a suitable replication scheme for replicating popular objects to well-performing nodes for increasing the availability and fault-tolerance. Due to the dynamic nature of nodes in an unstructured network, nodes may come up and down frequently. In order to provide efficient access to popular files for all nodes in the network, only popular objects are replicated to good nodes.

The migration policy of elephants influences the proposed resource discovery scheme. Each query is treated as an elephant. Each node houses a collection of objects. A group of peers hosting the resources within certain hop-limit forms an *area*. The area may be a *dry area or wet area* depending on the number of popular objects existing in the area. The service a node in the network may receive within a dry area is very limited as the area contains less number of popular objects. Hence queries have to travel long distances for successful resource discovery and most of the queries may fail. When the number of hits in the dry area reaches below a threshold, the queries are forwarded only through powerful peers situated in the wet area. The peers located in wet area are called power peers. The wet area hosts large number of popular objects, and at the same time a proper mechanism for moving popular objects from wet area to dry area is required. Nodes having several power peers as neighbors may receive more query hits since queries from power peers are only passed through other power peers. Hence, such nodes may be always situated in a wet area.

A power peer is a powerful node in the sense that it has high degree; large number of popular objects, and high bandwidth. A bunch of power peers form a wet area. These power peers together participate in the search process once a node does not receive adequate hits in a dry area. A node maintains a list of power peers in a data table. This table is called *power peer table*. During a search process, for each query hit, a querying node checks the category of node from which a success has occurred. If the node is a power peer, and its identity does not exist in the table, an entry for the power peer is created. Thus, the number of entries in the data table increases with number of query hits through different power peers in the network. Finally, a node may have a collection of power peers located in the network. The IDs of power peers along with their bandwidth, and degree are recorded in the table. The table also maintains data about number of hits in each power peer for the queries originated from the node.

For each neighbor, a node 'N' in the network maintains a variable which contains the number of hits through the neighbor for the queries submitted to it for a certain period of time 't'. The hit rate is also recorded. Hit rate is computed as the ratio between number of hits to number of queries inputted. The average of hit rate for all neighbors for time 't' is computed and if it is less than a threshold $(\delta)$, the node is said to be located in a dry area. After that, the node tries to move from dry area to wet area due to low hit rate it receives. The future queries are forwarded only through the power peers chosen from its power peer table. For that, the values of number of hits $(h_p)$ through the power peer for the queries originated from the node, degree $(d_p)$, and bandwidth $(b_p)$ are utilized. For each power peer in the table, a *utility value* $(u_s)$ is computed as in equation (1). K power peers with high utility values are chosen for routing K walker messages. The utility value for a power peer is updated based on the result of every query message forwarded to it. The utility value increases with high success rate, association with more neighbors, and high bandwidth.

$$u_s = (w_1 * h_p + w_2 * d_p + w_3 * b_p) * 100 \quad ........(1)$$

where, $w_1 + w_2 + w_3 = 1$, $w_1, w_2$ and $w_3$ are the assigned weights for normalizing the values.

A power peer forwards a query message to another power peer only. Thus, the queries are passed through a resource fertile wet area. The walkers are terminated when either result is found or TTL is exhausted. The processing of queries in a wet area may provide improved success rate with less search path length as the power peers hold more number of popular objects.

Shifting queries from dry area to wet increases the query load among power peers. Hence, appropriate measures are required to handle this situation. In the first mechanism, when CPU load on a power peer reaches certain threshold, an overloaded power peer forwards the incoming queries to other power peers with next higher utility values in the power peer table. The second mechanism is to increase the availability of popular objects in the dry area. This is done by means of a replication technique. Q-Erasure autonomously replicates objects to well-performing nodes in dry as well as wet areas. In the context of replication, a well-performing node is one which possesses high bandwidth, high degree and large available free storage. Hence, even if a node shows low performance in resource discovery, due to the presence of said parameters it is selected as a candidate node for hosting a replica. Due to replication a dry area gradually becomes green.

In the context of proposed search scheme, for successfully replicate objects to dry and wet areas a small variation of Q-Erasure (section V) is employed. When a node shifts to wet area, the neighbors of the node do not further directly receive any queries from the node. With the purpose of increasing availability of objects, neighbors possessing good values for bandwidth, degree and free storage are the good candidates for hosting new replicas. So, other power peers listed in the power peer table accommodate a few selected neighbors of the node N in their replication Q-table. The higher the values of bandwidth, degree, and the amount of free storage available in the shared folder of every neighbor $(f_p)$, higher the chances of getting accommodation in Q-table of a power peer. For this, every neighbor of the node transfers its bandwidth and free storage level to the node based on its request. The node computes another utility value $(u_p)$ as in equation (2). The neighbors of node N possessing utility values greater than a threshold are accommodated into the Q-table of power peers listed in the power peer table. We assume that every power peer permits an ordinary node to assign its well-performing neighbors in their replication Q-tables.

$$u_p = (w_1 * f_p + w_2 * d_p + w_3 * b_p) * 100 \ldots\ldots(2)$$

Next step is to assign the chosen neighbors to the power peers in the table. This is done by the node itself using the utility values of power peers - $u_s$. The node N computes the ratio $(n_p)$ between number of chosen neighbors and number of power peers in the power peer table to the nearest integer. This value represents the number of neighbors to be assigned to the Q-table of each power peer. The $u_s$ and $u_p$ values are sorted and the node N compares both values. Power peers having higher utility values of $u_s$ assigns $n_p$ neighbors with higher utility values for $u_p$ in its Q-table. The number of power peers or chosen neighbors may be odd or even. Hence, it is not mandatory that all power peers will receive the same number of neighbors of node N during this process. All the new members which are entered into the Q-table of each power peer receive an initial Q-value of 100. According to the Q-values, the new entrants receive replicas of popular objects from power peers. Thus, the availability of popular objects is augmented.

Once a dry area is crowded with sufficient number of popular objects, the area gradually becomes wet. Hence, the neighbors that are shifted to the power peers can attain the previous status to receive incoming queries from N. A few steps are to be done for completing this process. Two values are associated with this operation – an availability threshold value $(\lambda)$ and hit rate threshold value $(H_{thrld})$. The ratio $(n_{avbl})$ of number of files replicated to the neighbor of node N listed in the Q-table of each power peer $(N_{repl})$ and, the sum of number of files actually present at the time of shifting to Q-table of power peer and $(N_{repl})$ is computed. If this ratio is greater than or equal to the availability threshold, the node N is informed and it records the status of the neighbor. Thus, when $x\%$ (say 80%) of neighbors possess high availability, N starts sending its query requests to the neighbors. The neighbors for whom a green signal is obtained from a power peer in the power peer table are only considered for sending queries. The routing through power peers in the power peer table is temporarily suspended. Moreover, the neighbors of a node act as neighbors for other members in the network. Hence, they may also receive replicas from the same member nodes. These replicas are taken into account while computing the replication ratio. As said previously, the hit ratio is $(H_t)$ also computed for each neighbor for a certain period. The power peer which holds the neighbor in its Q-table periodically collects the hit ratio from N and if the hit rate is greater than or equal to $H_{thrld}$, the power peer assumes that node N is situated in a wet area. However, the power peer continues to serve the node by replicating popular contents to neighbors in the Q-table. But, the neighbors in the Q-table which are having a $n_{avbl}$ less than half of the threshold value up to the period are removed from the Q-table due to their low performance in replicating objects autonomously. This also saves computational resources of power peers. The major steps in the proposed search technique are portrayed as an algorithm (Algorithm 1).

*Algorithm 1: Searching*

If $(AvgHit\ Not < \delta)$ then
……..Node N Continues searching through its neighbors
else
Node is in dry area – DRY
Node in wet area – WET
Starts searching through selected power peers in power peer table
If query load on a power peer exceeds to certain level, redirects the incoming queries to suitable power peer in the power peer table based on its utility value
Power peers replicate popular objects to DRY area
If $(n_{avbl} \geq \lambda)$ for $x\%$ of neighbors of N then
    Node starts sending queries to neighbors in DRY area
    Suspend the sending of queries to power peers located in WET area (*Power peer Table*)
    Compute hit rate - $H_t$ for period 't'

    If $(H_t \geq H_{thrld})$ then
        Area DRY is declared as new WET area
        Node N continues searching through its neighbors as previously

## V. Q-Erasure

The proposed replication technique - *Q-erasure* utilizes the advantages of erasure coding and Q-learning [45] to enhance availability of popular objects in the network. In a P2P system, a file can have several erasure coded replicas. A node in the network maintains a few data structures. Q-table is one data structure which contains node-IDs, and corresponding Q-values. The Q-values are updated for each operation irrespective its success or failure.

A node hosts two kinds of resources: *full files and erasure coded blocks.* The full files are used by a user for meeting his/her requests, and replication to other sites. The full files are not sharable for user queries. Only the erasure coded blocks hosted in a node are shared. It is not mandatory that the erasure coded blocks should contain blocks of same files hosted in full file folder. It may contain blocks of other popular files in the network. The replicated erasure coded blocks are distributed to folders which hosts erasure blocks. The important attributes of files and blocks being hosted by a node are also maintained in a database. When the popularity of an object reaches certain threshold, the object is chosen for replication. Erasures coded blocks are created from a full file, and these blocks are replicated to other peers in the network based on their past performance.

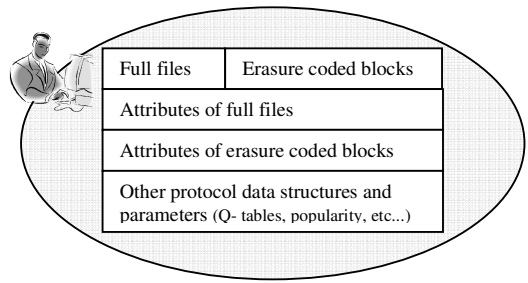

**Figure 2. Data structures (Q-Erasure)**

---

*Algorithm 2: Q-Erasure*

// S – List of target peers for hosting replicas total
// k – maximum number of blocks to be created by erasure coding
// $P_{th}$ – popularity threshold
// $H_{sh}$ – hash value of a full file 'F' having a erasure coded block B
// FULL – directory of full files in a node
// BLOCK – directory of erasure coded blocks for sharing in a node
// $E_r$ – An erasure coded block hosted in a node
// $Q_{th}$ – Minimum Q-value threshold (preset as 100)
// $B_f$ - Blocks already created for F (F is partially replicated file)
//Replication List- *a table that contains a list of object names (hash values) reserved by other nodes during the object checking process*

For F in BLOCK with popularity $P_f \geq P_{th}$
    Check whether a copy of F with $H_{sh}$ exists in FULL
    **If** F exists, select F that is replicated partially or not yet replicated
        Choose appropriate peers for hosting replicas from the Q-table
            Compute the average of Q-values -AvgQ
            Select the peers with $Q-value \geq AvgQ$ and $Q-value \geq Q_{th}$
        For each chosen peer, verify $H_{sh}$ of F exits in it's BLOCK
    If a peer is not up, or minimum two blocks of F exist in BLOCK or $H_{sh}$ exists in the *Replication List* leave out the peer from the replication process
   from the remaining chosen peers identify a list of target peers S for hosting replicas
   If F is a partially replicated file
      Compute number of blocks to be replicated for partially replicated file, $BP = k - B_f$
    If $(S \geq BP)$
      Replicate BP blocks to the same number of target peers
      Remaining each eligible peer receive one random replica from already generated blocks of F
    Else generate S blocks and replicate to chosen peers according to Q-values
   If F is a file which requires replication of k erasure coded blocks
      If $(S \leq k)$, generate S blocks randomly and assign each block to target peers according to their Q-values
    Declare F as a partially replicated file for $(S < k)$ condition
      Else if $(S > k)$
        Replicate one block of F each to k peers according to their Q-values
        Remaining eligible peers receive random replicas from already generated blocks of F
Update the status of F as replicated or not, number of blocks replicated hosts of replicated blocks etc…
For each replica a peer in the Q-table receives
    Gather the bandwidth, available free storage and degree of each peer which have received replicas.
    Compute reward for each peer in Q-table
$$\rho_i = \left[ \left( \left[ \frac{d_d}{d_{min} * w_1} \right] + \left[ \frac{b_w}{b_{min} * w_2} \right] + \left[ \frac{s_{avbl}}{s_{min} * w_3} \right] \right) * 100 \right]$$
    Update Q-values of each peer in the Q-table using computed reward and current Q-value
      Update the Q-value of node, which has received a replica as: $Q_{i,t+1} \leftarrow Q_{i,t} + \alpha(\rho_i - Q_{i,t})$

Nodes with a copy of the object, which are excluded in the replication process, do not alter their Q-value.

---

The performance of a peer is represented as Q-values in the Q-table of a node. The attributes of each object such as its name, unique hash value, a list of keywords for each object etc. are also maintained in a node. 'k' erasure coded blocks can be created for an object. But the number of blocks to be created depends on the attributes of participating peers accepting the replicas. Hence, not all blocks are created simultaneously if sufficient peers meeting the requirements are not available. After each replication process, the Q-tables are updated with new Q-values. Q-Erasure selects the desired object (full file) to create erasure coded blocks, selects appropriate sites, determines the number of blocks to be created at a time for an object, object eviction in case of storage exhaustion, and updates Q-values using various gathered parameters. The entire process of Q-erasure is depicted as an algorithm (*Algorithm 2*).

### A. Selection of Objects for Replication

The objects are chosen for replication according to their popularity. An object is decomposed into N fragments. K erasure coded blocks are generated from these N fragments. These blocks are distributed to various peers in a P2P network. The frequently accessed erasure coded blocks from the shared storage space of a node are treated as popular objects. Every erasure coded block is associated with its corresponding full file name, equivalent hash value and number of times it is accessed (downloaded by other peers in the network). The rank of an object for its erasure coded blocks hosted in a node relies on total number of downloads for all the blocks hosted in the shared folder for different objects - $Total_{down}(t)$ for the period 't', and number of times each block $E_{r_i}$ of an object having $N_{er}$ number of blocks hosted in a node are downloaded $(T_i)$ during 't'. The high value of rank indicates that the object is a most popular item. Each block hosted in a node holds meta-data such as object name, hash value of the object and so on.

The system regularly (e.g. for every 100 requests received by a node) modifies the popularities of all the blocks in the nodes. The popularity $P_f(t+1)$ is modified as in Eq. (3)

$$P_f(t+1) = \left( \ln\left(\frac{P_f(t)}{w_1}\right) + \eta \left[\frac{\sum_{i=1}^{N} T_i}{Total_{down}(t)}\right] \right) x10 \ldots\ldots(3)$$

In Eq. (3) the value of current popularity of the object $P_f(t) \geq 0$ and the value of constant $\eta$ is in between 0 and 1. The value of constant decides the level of contribution to the current popularity value. The higher the value of constant $\eta$, higher is the level of contribution to the popularity. If the current popularity value $P_f(t)$ is zero, the modified value of popularity is kept as zero minimum. The $w_1$ is a weight associated with the current value of popularity and it is in between 0 and 1.

The modified value of popularity is written into a table (*popularity table*) after removing the existing values. The update process also utilizes the existing popularity value. The initial value of $P_f$ for an object is always zero. If the number of downloads for all the blocks hosted for an object for the time period 't' is nil in a node, the popularity of the object is reduced to $\left[\ln\left(\frac{P_f(t)}{w_1}\right)\right]$ denoting low popularity of the object. The reduction in the value depends on the selection of $w_1$. Assigning large value to $w_1$ reduces the popularity of the object drastically; hence its value is kept as minimum in this work. Otherwise, the popularity of the object increases with the number of downloads of each of its block at the node.

**Table 1. A sample popularity table**

| Object Name | Hash value | Erasure coded blocks | | | Popularity |
|---|---|---|---|---|---|
| Ai.txt | e0d123e5f316bef7 8bfdf5a008837577 | $A_1$ | $A_2$ | $A_3$ | 0.4286 |
| | | 2 | 0 | 1 | |
| Bi.txt | 35d91262b3c3ec88 41b54169588c97f7 | $B_1$ | | $B_2$ | 0.5714 |
| | | 3 | | 1 | |
| Ci.txt | cc273fe9d442850f a18c31c88c823e07 | $C_1$ | $C_2$ | $C_3$ | 1.0000 |
| | | 4 | 2 | 1 | |
| Di.txt | 41b5416e0d123e5f 316bef78588c97f7 | $D_1$ | | | 0.0000 |
| | | 0 | | | |

A sample popularity table with important parameters is shown in Table 1. Three erasure coded blocks are hosted at the node for an object Ai.txt. The block $A_1$ is downloaded two times and that of $A_3$ one time during the period 't'. The initial value of $P_f(t)$ is set at zero. The total number of downloads for all the blocks in the node is 14. In this example $w_1$ is preset as 0.2. At the end of the period, the system computes the popularity using the above parameters as 0.4286. This value is updated for every period.

For every $\delta$ period $(\delta > t)$, the system identifies the possible candidates for replication. This is done by comparing the popularity of a block at the time of δ with a threshold popularity value $P_{th}$. When the popularity of a block at $\delta$ becomes greater than or equal to the threshold value, i.e. $P_f(\delta) \geq P_{th}$, system checks for an object with the same hash value as in the meta data of the block exists in the directory of full files (FULL) and if exists the object is chosen for replication.

### B. Q-table creation and Initialization

Once a popular object is selected for replication, next step is to locate suitable peers for hosting erasure coded blocks of the popular object. The target peers are chosen from a Q-table maintained in each node. The Q-table contains peer-IDs and respective Q-values for each peer. The high Q-value represents the soaring performance of a peer in the past. The members for the Q-table are assigned after a simple operation: a message (Hello message) is sent to peers that come within a time-to-live (TTL) limit, which is the number of hops the message should be propagated; the responded peers become members of Q-table with some initial Q-values. The message forwarding follows a k-random walk [8] procedure. Initially K messages are generated and the messages are propagated through K number of neighbors selected randomly. Neighboring nodes forward the message to one of their neighbors; from there to next hop. The message has a message-id. Nodes, which have already received a copy of the message, keep the message-id and address of the neighboring node to which the message was forwarded. Hence, when a node receives the same message another time, it will not be forwarded to a node that has received the message previously, however it selects a different peer from the neighbor list. The response messages from the peers consist of equivalent values for their current bandwidth $(b_w)$, and available storage $(S_{avbl})$. Using these values, Q-tables are initialized. The P2P system assigns minimum values for node attributes such as bandwidth $(b_{min})$ and storage $(S_{min})$, which are used for Q-value computation. The Q-values of each node in the table is initialized as $Qr = \left(\frac{b_w}{b_{min}} + \frac{s_{avbl}}{s_{min}}\right) x100$. In order to eliminate the random or probabilistic assignment, Q-values are thus initialized with important resources bandwidth and storage.

### C. Selection of nodes

In order to replicate the selected object to well-performing peers, the status of the object is checked. The object may be in three states: *fully replicated, partially replicated or not replicated.* If 'k' blocks are replicated successfully, the object is said to be fully replicated. In case, sufficient target peers are not available, not all blocks are replicated. Hence, the object is in partially replicated status. Even if, a fully replicated object still has high popularity value in the future, it is again replicated to other chosen peers, if available in the Q-table. The remaining blocks of partially replicated objects are replicated if adequate number of peers are available in the Q-table. The blocks of objects with not replicated status are replicated suitably based on availability of required number of target peers in the Q-table of the node.

A peer, which can host a replica, should have a high-speed connection, minimum available storage, link with more number of nodes (degree) and it should stay online for a longer period. From the possible set of host candidates listed in the Q-table, the best ones according to the bandwidth, available storage, and number of links (degree) are chosen. The objects are copied into nodes, which do not already host the same replica of the target file. Hence, the overwriting of the same file in a node is avoided and at the same time, the process saves bandwidth consumption due to redundant file transfer. In order to choose possible candidates for hosting the replica, the mean of Q-values listed in the Q-table is computed. Nodes with Q-values greater than or equal to the mean (AvgQ) are selected and a message is sent to each selected node to verify whether any block of the chosen object exists in their shared folder-BLOCK. *Replication List* of a node is a data structure that contains a list of object names or hash values of objects reserved by other nodes during the object checking process. This avoids other nodes to replicate the block of same object to a peer as the same peer may be chosen by another peer as a target peer for replication. If the node is not up, a copy of the object is present, or the object's name appears in the replication list, the node is left out from replicating the chosen file. All other nodes with Q-values greater than or equal to AvgQ are selected as target peers for hosting replicas.

### D. Replica Distribution

Replicas are distributed based on the status of an object and number of chosen peers. If object 'F' is a partially replicated file, the number of remaining blocks to be replicated 'BP' is computed from the maximum number of blocks 'k' for an object and blocks already replicated $B_f$. The large number of target peers (>BP) available for hosting new blocks permits to replicate blocks to more than BP number of peers. Randomly selected blocks from these BP blocks are thus distributed to the remaining peers.

For an object which requires full replication needs k target peers. At a time, only one block of an object is replicated to a peer. A peer is allowed to host a maximum of two blocks of an object out of its k blocks. Thus, the blocks of an object are distributed to several sites in the network. If sufficient peers are not available for hosting the blocks, system generates blocks randomly and assigns them to available peers. The object is declared as partially replicated. Q-learning modifies the Q-values of peers in the Q-table based on their performance. Hence, in the future, if the peers in the Q-table of a node perform well, their Q-values will increase and thereby participate in the replication process. At that time, the remaining blocks of partially replicated objects are copied into these well performing peers.

### E. Reward computation

The nodes, which have received an erasure coded block, send the values for degree $(d_d)$, bandwidth $(b_w)$ and available storage $(s_{avbl})$, to the node that initiated the replication process. This is the reinforcement signal to the replication system. Based on the reinforcement signal, the reward ($\rho_i$) is computed for each node in the Q-table as $\rho_i = \left[ \left( \left[ \frac{d_d}{d_{min} * w_1} \right] + \left[ \frac{b_w}{b_{min} * w_2} \right] + \left[ \frac{s_{avbl}}{s_{min} * w_3} \right] \right) * 100 \right]$, where $w_1 + w_2 + w_3 = 1$. As the bandwidth is a very important network resource, priority is given for it while computing the reward, hence $w_2 < w_1, w_3$. Therefore, the nodes with large bandwidth highly influence the reward. Moreover sufficient storage space should be available in a node for hosting more and more replicas of different objects. In a P2P network, a few nodes have a large number of degrees while most of other nodes have only a small number of degrees. Peers with a large number of degrees make many replicas as peers with a small number of degrees. In addition, replicas on large degree peers are used frequently as those on peers with small degrees. In our strategy, the system assigns a common minimum degree threshold ($d_{min}$) value to be used for replication to all nodes. In terms of *degree,* the contribution of high degree nodes to the reward is high as compared to low degree nodes. At the same time, nodes with only high bandwidth and storage can also participate in the replication process. All these factors ensure the availability of objects within short hop distances.

*F. Q-table Update*

The reward values are utilized to modify the Q-values. The update process increases, or decreases the Q-values of peers that are being participated in the replication process. The nodes, which have not participated, do not modify the present Q-values. The nodes with high Q-values are treated as good peers. The Q-values of nodes, which have created a replica, are updated as $Q_{i,t+1} \leftarrow Q_{i,t} + \alpha(\rho_i - Q_{i,t})$, where α is the learning rate (value of $\alpha$ between 0 and 1), and $Q_{i,t}$ is the present Q-value. If the reward of replication is high, the Q-value is incremented and it relies on bandwidth, available storage and degree of the node. The current Q-values are retained for the nodes comprising a copy of the object i.e. $Q_{i,t} \leftarrow Q_{i,t}$. Nodes that are not up are punished heavily with zero reward, $\rho_i = 0$ and the Q-values are updated as $Q_{i,t} \leftarrow Q_{i,t}(1-\alpha)$. Assigning high value to the learning rate constant yields a large increase in Q-values of nodes that have placed a replica to their respective directories.

*G. Object Replacement*

Some blocks which are hosted in a node should be deleted to make space for new blocks of popular objects if adequate storage space is not available in a node. Q-Erasure removes a block according to number of downloads and its age. The age attribute represents the time at which a block was inserted into BLOCK. If a block is recently added to the shared directory, it may have less number of downloads and small value for age. Hence, a block with less number of downloads and large value for age is removed for housing new erasure coded blocks if adequate space is not available. The addition of a new block creates new entries in the popularity table.

**VI. Experimental Setup**

The experimental setup is similar to the settings in [46]. The simulation tool is developed using Java language. The tool runs in a Windows operating system environment. The software, which are used for developing the simulation software are NetBeans, J2SE Development Kit and WampServer. NetBeans is a free, open-source Integrated Development Environment, which supports development of all Java application types. WampServer is an open source project and Windows web development environment. It allows creating various applications with Apache, PHP and the MySQL database. WampServer also includes PHPMyAdmin and SQLiteManager for managing databases.

The proposed techniques are simulated using random graphs that have 10000 nodes. The nodes can join the network and establish random connections to existing nodes. Each node carries a few files. The average degree of a node in the network is 3.5. There are 1000 objects distributed to various nodes, the objects are replicated to different sites using Q-Erasure. The quantities of objects maintained in the system are sufficient to analyze the performance since Q-Erasure effectively propagates the erasure coded blocks of popular objects to various sites in the network. The objects are Microsoft word, PDF and text files available as course materials on various subjects such as computer science, electronics, physics, mechanics, electrical etc. Thirty thousand keywords are chosen from the course material files and these words are randomly selected as query keywords by all the nodes during searching. Two types of query searches are employed: file name based and keyword-based. In the file name based search only the objects names in the shared storage space of each node is searched. The objects containing the keywords are looked up in the keyword-based search. In the simulation scenario, all the queries contain keywords alone.

Each node generates 100 queries and one query is propagated every 20 seconds on average. However, each node enters the query generation phase in a randomly selected time slot. Hence, the flood of query message production is regulated. 80% of the nodes are up at the time of performing simulation. 50% of 'Down' nodes selected randomly change their status to 'UP' after every 50,000 queries are propagated and, at the same time, the same amounts of UP nodes obtain the DOWN status. The conditions that are employed for forming dry and wet areas are not employed to avoid complexity. Based on hit rate, a node moves to wet region. The hit rate threshold δ is preset as 0.3 and the value of $H_{thrld}$ is 0.6. Availability threshold $\lambda$ is fixed as 0.4. The maximum number of blocks k to be created for an object is preset as 12. There will be 8 fragments for each object.

The default TTL value is preset as 6. There are ordinary peers and power peers. The power peers are selected based on node their degree, number of objects being hosted and free storage. A node with its degree greater than or equal to 7, available storage ≥ 30% of the shared storage space and number of shared objects in a node≥ 15 is declared as power peers. The minimum degree of a power peer is preset as seven. Initially twenty percent of the nodes in the network are assigned the power peer status. The maximum cpu_load that can be processed is represented as number of messages a node can handle at a time. This value remains the same for each power peer until the end of simulation. The average degree of a node in the network is 3.5.

## VII. Results and Discussion

Experiments are conducted several times with the objective to measure query success rate and messages per query.

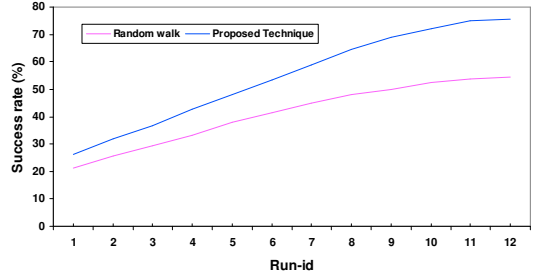

**Figure 3. Query hit rate for random walk and proposed scheme**

*Query Success rate- random walk vs. Proposed scheme:* The aim of conducting this experiment is to measure the success rate for queries generated by different nodes in the network. The experiments for the proposed technique are conducted in the same environment as above; however, for random walk a slightly different environment is used. For testing the performance of random walk search technique, no means of replication is employed. Moreover, all nodes have equal status. Experiments are conducted several times and the results are plotted as a graph shown in figure 3. From the graph we can conclude that the proposed search technique performs far better than random walk in terms of success rate. The success rate of the proposed scheme increases with time and some times later the rate of success go steadily. This is due to the effective distribution of erasure coded blocks of popular objects to well-performing nodes by Q-Erasure. So, more regions gradually become fertile causing increased success rate.

*Average % of query success rate through ordinary nodes and power peers:* Aim of this experiment is to measure contribution of neighbors' of peers and other power peers in the network towards query success rate. Initially, most of the queries are failed and slowly this situation changes. As more nodes are moved to wet area, the successes through power peers are increased. This is implicit from the results shown in figure 4 for run-IDs 2 to 6. From seventh simulation run onwards, neighbors play important role in increasing the query hits. This keeps on increasing; finally the major contributors are neighbors. The proposed scheme moves the neighbors in the dry areas to wet areas and thus due to replication these nodes receive more amount of popular objects. Moreover, the amount of failed queries is significantly reduced. Because all categories of nodes are contributing to the increased success rate, query load among nodes are properly balanced.

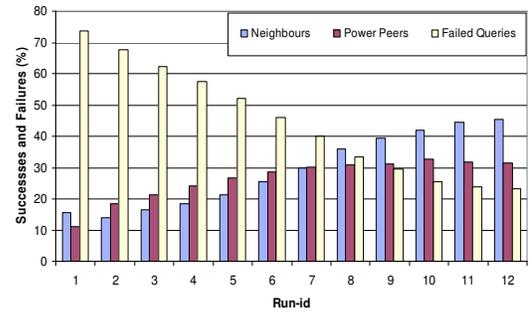

**Figure 4. Success rate through different types of nodes**

*Messages per Query:* To study how the proposed scheme affects the network traffic, the amount of messages generated for each query is monitored. Random walk produces approximately the same quantity of messages for each query being generated (figure 5). However, the proposed scheme initially takes more messages per query and it gradually decreases as time goes. This is due to the high availability of blocks of popular objects in near by areas and large switching of dry areas into wet areas.

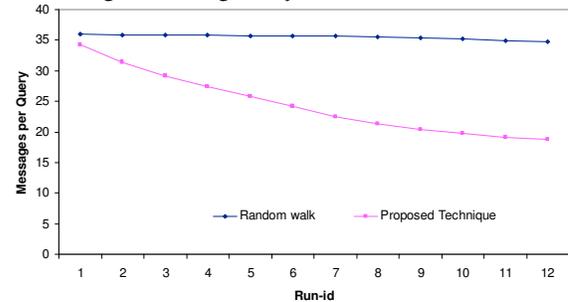

**Figure 5. Messages per query**

*Queries finished (k-random walk with path replication):* simulation experiments are conducted in a random network comprising 10000 nodes to compare the performance of path replication [8] and Q-Erasure on random K-walk search technique [8]. Path replication replicates an object along the path of a successful "walk". It does not cover any other node in the network for hosting replicas. The number of walkers are limited to 6. The results for queries finished in each simulation run are shown in figure 6. The success rate of random k-walk search with Q-Erasure is higher than the success rate produced for random k-walk with path replication technique. The influence of Q-Erasure in success rate improvement is very high as compared to path replication in each simulation interval. Each simulation run creates new replicas of popular objects in various nodes. Q-Erasure creates replicas for blocks of popular objects in more number of nodes. On the other hand, path replication creates replicas on nodes which are on the same search path. Also, in path replication, the target nodes to host full file replicas are not selected based on their performance in the past.

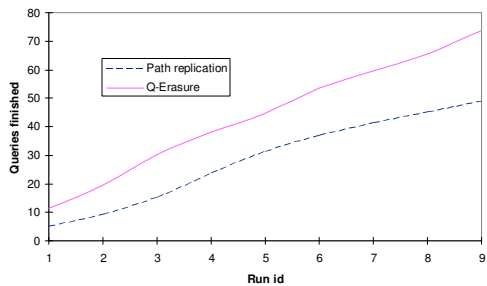

Figure 6. Percentage of queries finished

## VIII. Conclusions

In this paper, a bio-inspired resource discovery scheme and Q-learning based erasure coded replication scheme for unstructured P2P network are proposed. The resource discovery scheme utilizes the migration policy of elephants in search of food during dry seasons. Each node in the network is classified as ordinary peers and power peers. A node maintains a list of neighboring nodes as well as power peers for routing queries. If enough resources are not available, the queries are propagated through the power peers located in the resource fertile area. At the same time, dry areas are filled with popular resources by means of erasure coded replication. Nodes possessing certain features are only chosen for hosting the blocks of popular objects. A variety of simulation experiments are conducted and the results show that the proposed technique significantly increases the query success rate and creates less network traffic. The replication technique effectively replicates popular objects to well-performing peers in the network.